\documentclass[aps,pra,twocolumn]{revtex4}

\usepackage{ae}
\usepackage{epsfig}
\usepackage{color}
\usepackage{dcolumn}
\usepackage{graphicx}
\usepackage{graphics}
\usepackage{amsfonts}
\usepackage{amsmath}
\usepackage{amssymb}
\usepackage{xspace}

\newcommand{\eq}[1]{Eq.\,(\ref{#1})\xspace}

\begin{document}

\preprint{2}

\title{Observation of shock waves in a large Bose-Einstein condensate}

\author{R.~Meppelink}
\author{S.~B.~Koller}
\author{J.~M.~Vogels}
\author{P.~van der Straten}
\affiliation{Atom Optics and Ultrafast Dynamics, Utrecht University, 3508 TA Utrecht, The
Netherlands}
\author{E.~D.~van Ooijen}
\email[]{ooijen@physics.uq.edu.au}

\author{N.~R.~Heckenberg}
\author{H.~Rubinsztein-Dunlop}
\affiliation{The University of Queensland, School of Mathematics and Physics, Qld 4072, Australia}
\author{S.~A.~Haine}
\author{M.~J.~Davis}
\affiliation{The University of Queensland, School of Mathematics and Physics, ARC Centre of Excellence for Quantum-Atom Optics, Qld 4072, Australia}

\date{\today}

\begin{abstract}
We observe the formation of shock waves in a Bose-Einstein condensate containing a large number of
sodium atoms. The shock wave is initiated with a repulsive, blue-detuned light barrier,
intersecting the BEC, after which two shock fronts appear. We observe breaking of these waves when
the size of these waves approaches the healing length of the condensate. At this time, the wave
front splits into two parts and clear fringes appear. The experiment is modeled using an effective
1D Gross-Pitaevskii-like equation and gives excellent quantitative agreement with the experiment,
even though matter waves with wavelengths two orders of magnitude smaller than the healing length
are present. In these experiments, no significant heating or particle loss is observed.
\end{abstract}


\maketitle

\section{Introduction}

The realization of Bose-Einstein condensates (BECs) in dilute atomic gases \cite{anderson} provides
the opportunity for the study of  non-linear matter wave dynamics.  Many experiments on both the
statics and dynamics of BECs have shown that experiments can often be modeled accurately by solving
the mean-field Gross-Pitaevskii equation (GPE) \cite{gpe, edwards, burger, trippenbach, dutton, truscott, katz, robins_rf1, robins_rf2, kohl, lye, gpe_interferometry}.  For example it has successfully been
used to model experiments on interferometry \cite{gpe_interferometry, gpe}, soliton formation \cite{burger, truscott}, four wave mixing \cite{trippenbach}, atom laser outcoupling \cite{robins_rf1, robins_rf2, kohl}, sound propagation and
superfluidity \cite{gpe} to name a few.  However, the Gross-Pitaevskii equation is not an exact
description of a Bose-Einstein condensate, but instead it is expected to be a good approximation
for condensates that contain a relatively large number of particles and are not undergoing violent
dynamics \cite{pethick, norrie}.  Thus it is of interest to experimentally probe the limits of
validity of the GPE in describing the dynamics of  Bose-Einstein condensates.

The well-known `Bosenova' experiment of Donley \emph{et al.}~\cite{Donley01} is one situation where
the validity of the GPE might be questioned, combining a small number of atoms with violent
dynamics.  The experiment began with a near ideal $^{85}$Rb condensate of a few thousand atoms in
its ground state before the atomic scattering length was manipulated using a Feshbach resonance to
being attractive.  The BEC was observed to collapse, emitting high energy jets and bursts of atoms.
A number of computational studies have shown good qualitative agreement with the experimental
observations \cite{Saito02,Adhikari02,Santos02}.  However, careful quantitative studies have
indicated that at the most basic level there is quantitative disagreement with the experimental
data.  Savage \emph{et al.} showed that the experimentally measured collapse time is typically
25\% \emph{shorter} than predicted by simulations of the GPE~\cite{Savage03}.  Going beyond
mean-field theory and including the first order quantum corrections to the dynamics was shown to
make little difference to the numerical results~\cite{Wuester05,Wuester07}.  A second group of
experiments by the same group found evidence for the formation of repulsive bright solitary waves
in the condensate collapse~\cite{Cornish06}.  However, modelling this experiment with the GPE is
unable to reproduce this experimental finding~\cite{Beata08}.

Another example where the GPE has successfully modeled results of violent experiments is in the
formation of quasi-1D bright solitons as reported in Refs.~\cite{Strecker02,Khay02}. Both of these
experiments involve a sudden change to an attractive scattering length. It has been shown
\cite{truscott} that the GPE can be used to successfully model these experiments. However, in this
case, the three-body re-combination, which is not included in the GPE derivation, plays a
nontrivial role in the dynamics, and the GPE must be modified by the addition of phenomenological
damping terms in order to agree with these experiments.

Another situation exhibiting violent dynamics in the solution of the GPE without dissipation is in
the generation of shock waves.  Damski \cite{damski} has calculated the 1D GPE dynamics following
the introduction and sudden removal of an attractive dimple potential in the centre of a
harmonically trapped elongated BEC.  The localised density bulge splits into two pulses, which
propagate towards the ends of the condensate.  Due to the density dependent speed of sound in the
system, the center of the pulse catches up to the front, creating a shock, and at this point the
calculations develop a strong fringe pattern with a spacing of the order of the healing length
corresponding to classical wave breaking. However, Damski speculated that in a physical system the
Gross-Pitaevskii equation would become invalid at this point and this would not be observed in an
experiment.

Recently, a number of experiments have been performed that have observed phenomena related to wave
breaking in a Bose-Einstein condensate.  The first experiments were by Dutton \emph{et
al.}~\cite{dutton}. They used the slow light mechanism to create a defect in the condensate which
was much narrower than the healing length. This defect created dark solitons which shed high
frequencies traveling at different velocities. The wave front of the propagating solitons
eventually became curled and decayed into vortex pairs.

Simula \emph{et al.}~\cite{Simula05} blasted a hole in a rapidly rotating oblate BEC using a
repulsive dipole potential and found qualitative agreement between their experimental observations
and numerical calculations of the GPE.  This was followed by Hoefer \emph{et al.} who performed
similar experiments on a stationary and slowly rotating oblate BEC, and made the connection to
dispersive shock-waves in the GPE~\cite{hoefer}.  They found that this wave breaking phenomenon can
be described by the Gross-Pitaevskii equation \cite{hoefer2}. However, to obtain quantitative
agreement between the simulations and the experimental observations they found that they had to use
the laser width as a fitting parameter.  In one situation agreement was obtained with a value 1.5
times larger than in the experiment, but in another it was required to be half as large, suggesting
that this is not a systematic error but something more fundamental.

The purpose of this paper is to generate shock waves in a large number, elongated BEC and to
perform a quantitative comparison with simulations of an effective 1D GPE as a test of its validity
in extreme conditions.  We observe no indication of heating or particle loss, and conclude that
even at these relatively high energy scales the GPE is valid for the time of the experiment.  Soon
after the experimental data for this work was taken, related work was reported by Chang \emph{et
al.} \cite{Chang08}. They also found good agreement with the GPE. However, these experiments were performed at significantly lower energy scales, where agreement with the GPE is expected, and hence they did not perform a detailed investigation of the fringe spacing and possibility of heating.

\section{Experiment}

In our experiment we begin with an almost pure sodium BEC containing up to 100 million atoms in the
$|F, m_F \rangle = |1,-1 \rangle$ hyperfine state. The atoms are magnetically confined in a clover
leaf trap with harmonic trapping frequencies of $97$ Hz in the radial direction and $3.9$ Hz in the
longitudinal direction \cite{stam}. For the condensate, this corresponds to a chemical potential of
214 nK and Thomas-Fermi widths of $20$ $\mu$m and $506$ $\mu$m in the radial and axial directions,
respectively. To create the initial disturbance we turn on a repulsive, blue-detuned laser beam
focussed in the middle of the BEC, intersecting the cloud in the longitudinal direction. The focus
has a waist of $90 \pm 11$ $\mu$m (1/$e^2$) and a wavelength $\lambda_L$, tunable from 567 to 584
nm, where the power can be switched within 10 $\mu$s. At $t=0$ the blue-detuned laser is turned on
suddenly with the magnetic trap still on and after variable times the cloud is imaged in the trap.
Note that the waist of the laser focus is much smaller than the axial width of the condensate and
much larger than the radial widths, suggesting that most of the dynamics will occur along the axial dimension.	

\subsection{Case 1}

For the first set of experiments we make use of phase-contrast in-trap imaging \cite{andrews}.  The
detuning of the imaging laser is chosen such that a $2\pi$ phase shift is accumulated for the
largest atom cloud density.  The resolution is $3 \times 3\mu$m and the images are recorded with a
CCD camera (Apogee AP1E).
\begin{figure}[h]
  \centering
    \includegraphics[width=0.75\columnwidth]{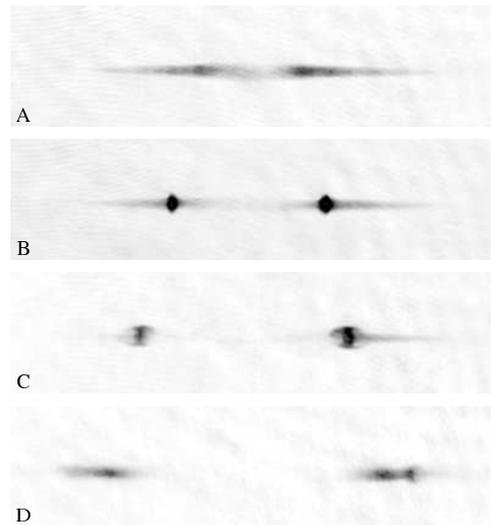}
\caption{Phase contrast images of the BEC in the trap for (A-D) $t=2$, 5, 8 and 15
ms after turning on the blue-detuned laser beam. The image size is 1200 $\times$ 300 $\mu$m.}
\label{insitu}
\end{figure}
Figure \ref{insitu} shows the in-trap CCD images for a BEC with 50 million atoms, corresponding to
a chemical potential $\mu= 162$ nK, imaged after $t= 2,6,8$ and 15 ms respectively. The power for
the repulsive laser beam used is 78 mW with a wavelength of 579 nm, resulting in a repulsive
barrier of 12.6 $\mu$K. For all experiments the scattering of photons from the laser  can be
neglected. The first image shows the splitting of the cloud induced by the switch on of the
repulsive barrier. The density profile of  the BEC results in a gradient in the speed of sound of
the condensate $v_c$ as $v_c \thicksim \sqrt{n}$.  As a consequence, as the wave fronts induced by
the lasers travel from the centre towards the edge of the condensate width the back of the pulse
catches up to the front, resulting in steepening of the wave (self-steepening).  This is clearly
observed in the second image. The last two images show that the wave breaks into two parts after
the maximum density gradient has been reached.

\subsection{Case 2}

\begin{figure}[h]
  \centering
    \includegraphics[width=1.0\columnwidth]{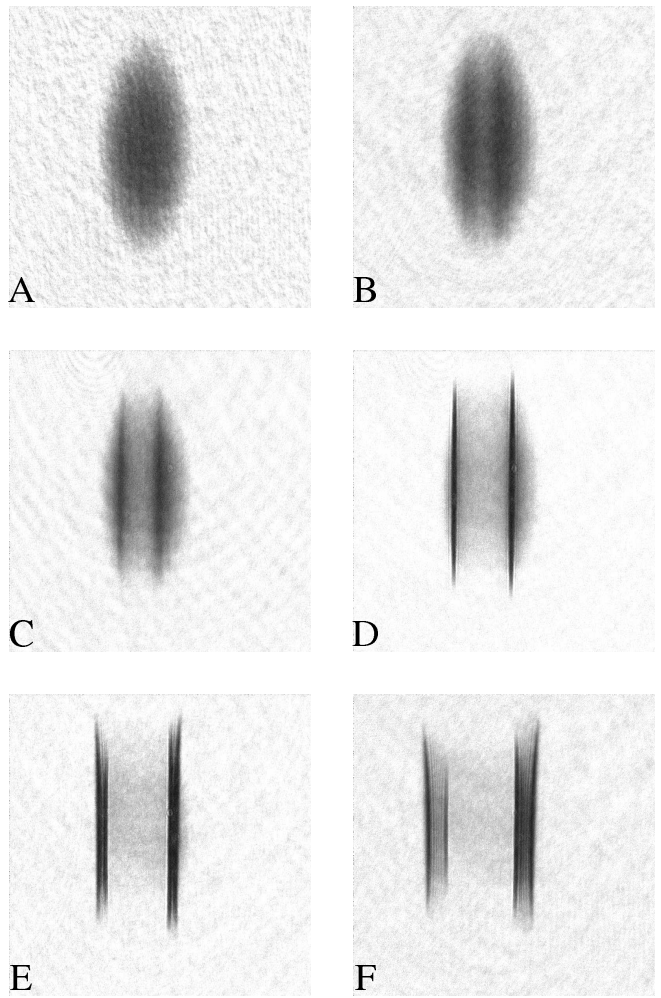}
\caption{Absorption images of an 70 ms expanded BEC for (A-F) $t=0$, 1, 2, 6, 13 and 25 ms after
turning on the repulsive barrier. The size of the images is 2304 $\times$ 1536 $\mu$m.}
\label{expanded}
\end{figure}

As the size of the fringes in the data set above was much smaller than the resolution of the
imaging system, the experiment was repeated and imaged using a 70 ms time of flight expansion to
enlarge the detailed features of the condensate. This free expansion time was chosen in such a way
that the full CCD array is used to obtain the maximum resolution. Figure \ref{expanded} shows the
images for a smaller condensate of 18 million atoms for an evolution time in the trap of $0, 1, 2,
6, 13$ and $25$ ms respectively.  In this situation we use a repulsive laser beam with a power of
$21$ mW and a wavelength of 569 nm, corresponding to a repulsive potential barrier with a height of
0.43 $\mu$K.

The images in Fig. \ref{expanded} indicated that it takes longer for the build-up of the wave front
for a smaller BEC and a weaker repulsive barrier. It can be seen that the shock front splits into
two parts, where pronounced fringe patterns appear. Fourier analysis on the last image shows a
fringe size of $6 \pm 1 \mu$m.  Integration over the entire pixel array shows a $2.7 \%$
fluctuation in the number of atoms over a evolution time of 30 ms, suggesting that losses due to
any potential heating are negligible. Frames (E) and (F) show that there is a slight curvature of
the wave fronts, suggesting that the outer edge of the wave front has travelled faster than the
center. 

\section{Numerical simulations}
We now turn to numerical modelling these experiments. Ideally we would do so with a fully
three-dimensional simulation of the Gross-Pitaevskii equation.  However, even making use of the
cylindrical symmetry we find that the simulation parameters for such a large BEC are too demanding
for numerical simulation with the computational resources we have available. To proceed we have
made use of the fact that the system has cylindrical symmetry and a high aspect ratio, and reduced
the problem to simulating the one-dimensional nonpolynomial Gross-Pitaevskii equation as derived by
Salasnich \textit{et al.} \cite{salasnich} (NPGPE).  This assumes a variational gaussian profile
for the wave function in the radial direction whose width, $\sigma(z)$, is dependent on the local
one dimensional density. This is to allow for the `bulging' in the radial direction due to the mean
field interaction to be included. The result is a 1D  Gross-Pitaevskii equation where the nonlinear
interaction term is a function of the local density. In the case of a harmonic trapping potential
in the radial direction with trapping frequency $\omega_{\perp}$, the effective 1D equation of
motion is
\begin{eqnarray}
i\hbar \frac{d}{dt} \psi(z) &=& \left( \frac{-\hbar^2}{2m}\frac{\partial^2}{\partial z^2} + V(z) + \frac{U_{3d}}{2\pi \sigma^2}|\psi(z)|^2 \right.\nonumber \\
&+&\left. \left(\frac{\hbar^2}{2m}\sigma^{-2} + \frac{m\omega_{\perp}^2}{2}\sigma^2\right)\right)\psi(z) \, , \label{npgpe}
\end{eqnarray}
with
\begin{equation}
\sigma^2(z) = \frac{\hbar}{m\omega_{\perp}}\sqrt{1 + 2a_s |\psi(z)|^2}\, . \label{sigma}
\end{equation}
$V(z)$ represents the external potential along the $z$ direction, $U_{3d} = 4\pi\hbar^2 a_s/m$ is
the 3-d atom-atom interaction strength, and $a_s$ is the $s$-wave scattering length. We solve this
equation numerically with no free parameters.

\subsection{Case 1}
We used \eq{npgpe} and \eq{sigma} to simulate the experimental conditions used in case 1.
We used an initial condition of $50$ million atoms in the ground state of a harmonic trap of trapping frequencies $\omega_z = 2\pi\times 3.9$ Hz and $\omega_{\perp} = 2\pi\times 97$ Hz. The ground state was found by imaginary time propagation. A repulsive potential due to the laser beam was then switched on. The laser power was $49$ mW, and detuned by $23$ nm from resonance. We assumed that the beam profile was gaussian with a width of $90$ $\mu$m. This corresponds to a repulsive potential height of $0.87$ $\mu$K.

\begin{figure}[h]
  \centering
    \includegraphics[width=1.0\columnwidth]{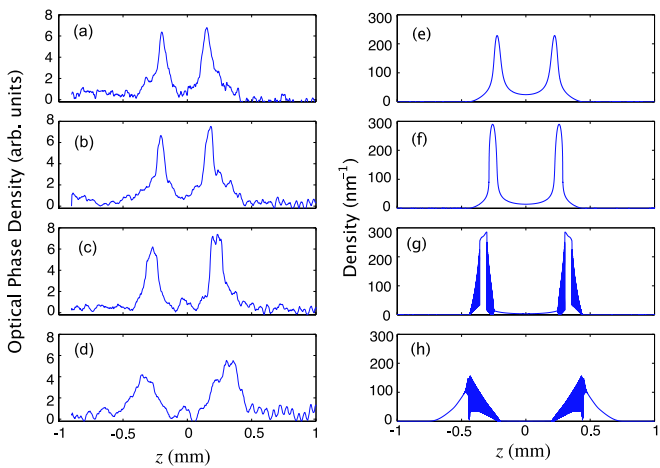}
  \caption{Comparison of theoretical calculations with experimental data. The left column ((a)-(d)) shows the
density profile of the condensate, integrated along the radial direction, and the right column ((e)-(h))
shows the theoretical calculations. The images are for (a) and (e): $11$ ms, (b) and (f): $14$ ms, (c) and (g): $21$ ms  (d) and (h): $32$ ms after the repulsive potential was switched on.  The laser power was $49$ mW, which corresponds to a repulsive potential height of $0.87$ $\mu$K} \label{trap_compare}
\end{figure}

 Figure \ref{trap_compare} shows a comparison of theoretical calculations with experimental data for these parameters.
The left column (a-d) shows the density profile of the condensate, integrated along the radial direction, and
the right column (e-h) shows the theoretical calculations. The images are for (a) and (e) $11$ ms, (b) and (f) $14$ ms,
(c) and (g)$21$ ms (d) and (h) $32$ ms after the repulsive potential was switched on.

\begin{figure}[h]
  \centering
    \includegraphics[width=1.0\columnwidth]{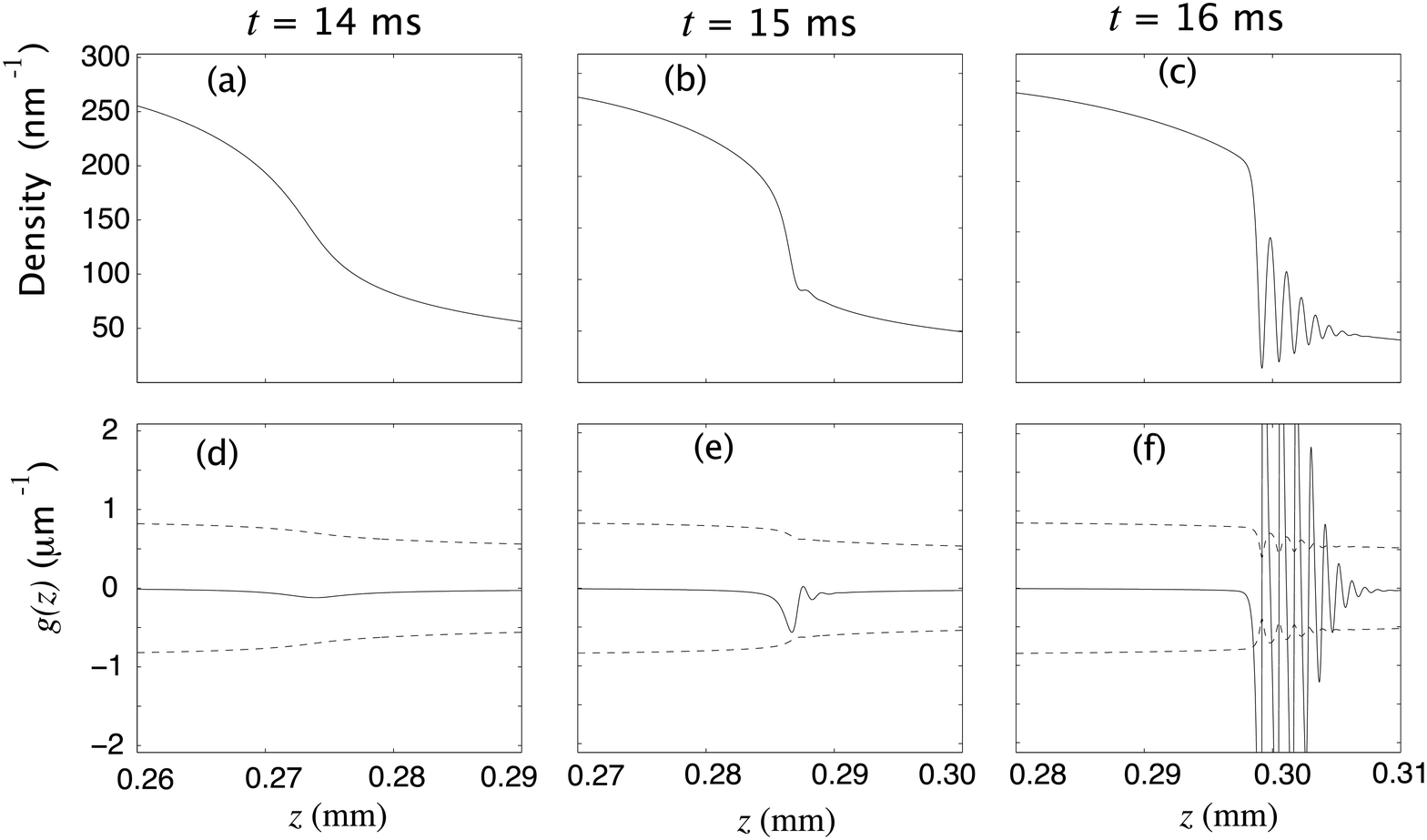}
  \caption{ Top row: Close up of the density profile of the leading edge of the pulse for (a): $t = 14$ ms, (b): $t=15$ ms, and (c): $t=16$ ms, for the same parameters as \ref{trap_compare}. Bottom row: The relative density gradient,
$g(z)$ (solid line) and the inverse of the healing length $\pm k_{wb}(z)$ (dashed line) for the
cases in (a), (b), and (c) respectively. In (d), $|g(z)| \ll k_{wb}(z)$, so the pulse propagates
smoothly. In (e), $g(z)$ becomes comparable to $k_{wb}(z)$, and a shock front begins to form. In
(f), $|g(z)| \gg k_{wb}$, and high frequency oscillations are visible in the density profile.}
\label{healing_length}
\end{figure}

The experimental images show that we get two pulses propagating outwards. The leading edges of these pulses gets steeper
 due to  the density-dependent speed of sound \cite{damski}. The theoretical plots show agreement with the experimental data in terms of the position of the pulses and the steepening of the density. However, they show extra detail which the experimental
image is insufficient to resolve. As the pulses steepen the relative density gradient become
comparable to the healing length, and a shock wave forms. This  causes an abrupt change in the density
profile of the smooth pulse, resulting in rapid density oscillations in front and behind the pulse.
Figure (\ref{healing_length}) compares the relative density gradient
\begin{equation}
g(z) \equiv \frac{\frac{d n(z)}{dz}}{n(z)},
\end{equation}
to the inverse of the healing length
\begin{equation}
k_{wb}(z) \equiv \frac{1}{\xi} = \frac{\sqrt{2 m n(z) U(z)}}{\hbar},
\end{equation}
where $U(z) = U_{3d}/2\pi\sigma^2(z)$ is the effective $1$-D interaction strength. In (a) and (d), $|g(z)| \ll k_{wb}(z)$, and the density
profile of the pulse remains smooth. In (b) and (e), $|g(z)|$ becomes comparable to $k_{wb}(z)$, and a shock front begins to form. In (c) and (f), $|g(z)| \gg k_{wb}$, and high frequency oscillations are visible in the density profile.

We found similar agreement between theoretical results and experimental data for repulsive barrier heights of $0.39$ $\mu$K and
$3.27$ $\mu$K (corresponding to data presented in figure (\ref{insitu})), which we have not presented here.

\subsection{Case 2}

\begin{figure}[h]
  \centering
    \includegraphics[width=1.0\columnwidth]{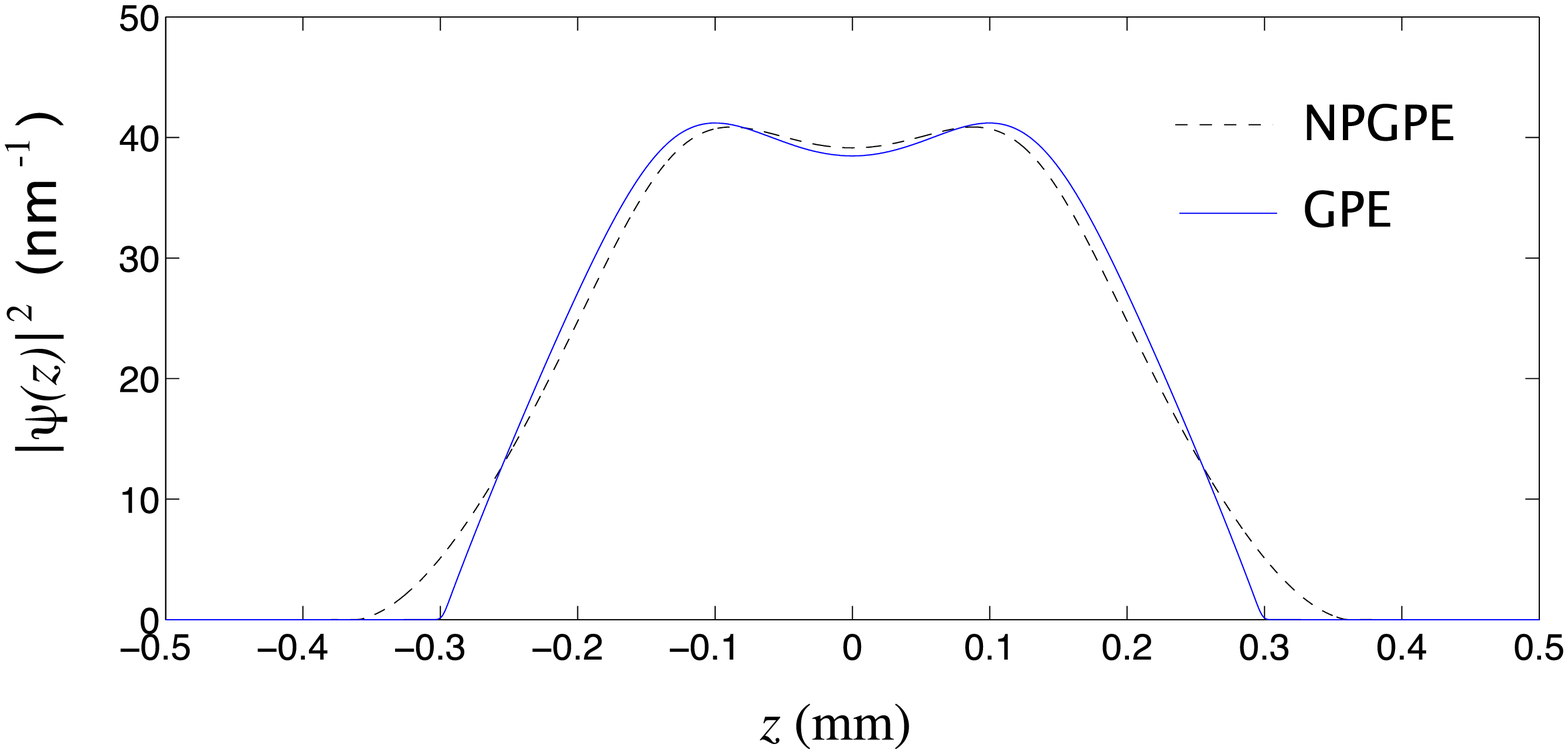}
  \caption{Comparison of the density profiles resulting from GPE (blue solid line) and NPGPE (dashed black line) simulations. The condensate was left to evolve in trap for $3$ ms with a $0.88$ $\mu$K repulsive barrier.} \label{np_compare}
\end{figure}

In the second set of data we made use of time-of-flight imaging and were able to observe features that were not resolvable in trap.  To make a comparison with this experimental data, we have modeled the expansion of the condensate in the radial direction during time of flight combined with continuing the 1D simulation of the dynamics of the wave function in the axial direction. We are unable to use the NPGPE in this case, as it is unable to model the expansion in the radial direction. Instead, we obtained a 1-D set of equations by assuming a transverse radial width of the condensate which was independent of $z$, which allowed us to scale our nonlinearity to one dimension using $U_{1d} = U_{3d}/\pi R^2$. We chose this width parameter $R$ by comparing the evolution while still in the trap to the evolution of the NPGPE with no free parameters, and chose the transverse width parameter which gave the best agreement in the results. Figure (\ref{np_compare}) shows the comparison between the solution to the GPE and the NPGPE after $3$ ms of evolution, during the trapped phase of the evolution. For the GPE simulation, we set $R=12$ $\mu$m, which gave the best agreement with the NPGPE.

 During the expansion phase, we assumed that the transverse width expanded according to the analytic result derived by Castin and Dum \cite{castin_dum} for the self-similar expansion of a Thomas-Fermi condensate. According to this result, the transverse width evolves according to $R(t) = R(0)\sqrt{1 + (\omega_{\perp} t)^2}$. As we have neglected to include three dimensional effects in our simulation, we do not expect perfect agreement with the experimental data.

\begin{figure}[h]
  \centering
    \includegraphics[width=1.0\columnwidth]{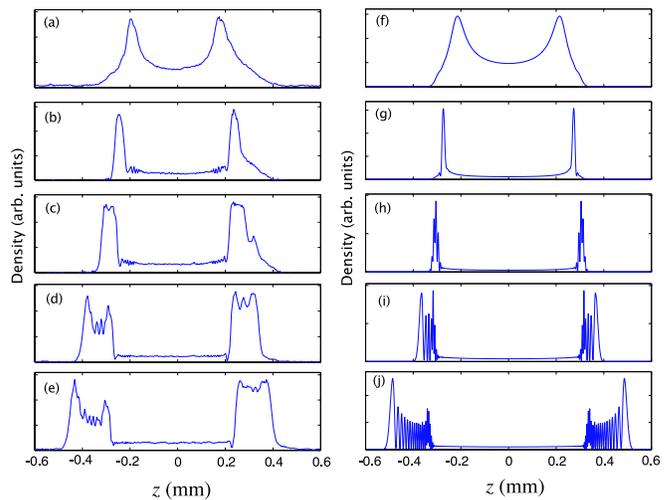}
  \caption{Density slice (left column), and theoretical 1-D GPE simulation (right column) of the
density profile of the condensate after expansion for $69$ ms after evolving in the trap for
various times. The condensate was held in the trap with the optical dipole on potential for (a) and (f):
$0.5$ ms, (b) and (g): $1.1$ ms, (c) and (h): $1.5$ ms, (d) and (i):$2.0$ ms, (e) and (j):$3.0$ms. Parameters: $N =
1.8\times10^7$, waist = $90$ $\mu$m, $P = 21$ mW, detuning = $10$ nm, corresponding to a barrier height of $0.88$ $\mu$ K.  } \label{expand_compare}
\end{figure}

Figure \ref{expand_compare} shows the a density slice from the experimental data from Figure
\ref{expanded} (integrated in the $y$ direction for $100$ pixels) compared to theoretical
simulations. There is excellent agreement between these results. In both cases, high contrast
fringes are observed when the back edge of the pulse catches up with the front edge.
We find that the wave breaking occurs after a much shorter time compared to the case I, because the expansion in the radial direction leads to a rapid increase in the
healing length.

\begin{figure}[h]
  \centering
    \includegraphics[width=0.8\columnwidth]{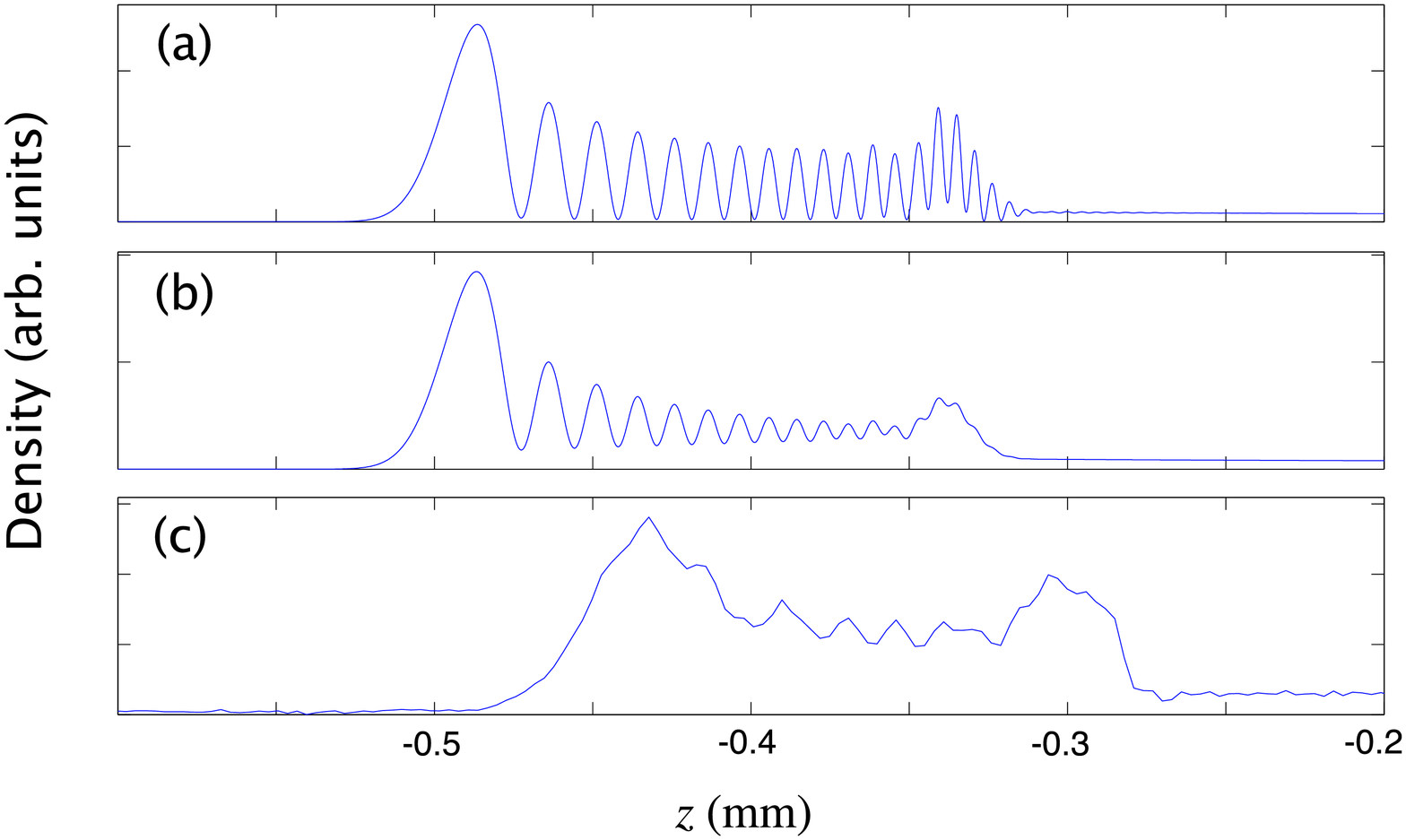}
  \caption{Close up view of the density slice for the same conditions as Figure \ref{expand_compare} (e) and (j). (a) shows the raw simulation data,
(b) shows the same data convolved with a $3$ $\mu$ m gaussian to emulate the finite pixel size of
the camera, and (c) shows the experimental data. The
discrepancy in the positions of the experimental and theoretical density distributions is most likely due to the uncertainty in the waist of the blue-detuned beam.   } \label{convfig}
\end{figure}

There is considerable agreement between our simulation with no free parameters, and the experimental data, indicating that it is valid to simulate such a violent system with the GPE. The main cause of discrepancy between the experimental images and the theoretical calculations is most likely the uncertainty in the size of the waist of the blue detuned beam ($90 \pm 11$
$\mu$m).  By adjusting the value of the waist slightly in the calculations, we found better agreement with the experimental images,
with the positions and widths of the wave packets giving best agreement for a waist of $92$ $\mu$m. There is also a slight discrepancy in the spatial frequency of the interference fringes for the experimental images and the simulation, the cause of which is unknown. It is difficult to accurately determine the spatial frequency in the experimental images, due to the $3$ $\mu$m pixel size causing spatial aliasing. The drop in observed fringe contrast in the experimental images is partly due to this spatial aliasing
effect, but most likely is the result of a slight misalignment of the imaging axis with respect to
the axis of the fringes in the 3-D experiment. The asymmetry is due to a slight misalignment of the
dipole barrier with the centre of the trap. Figure (\ref{convfig}) shows a comparison between the experimental data, and the results from the simulation convolved with a gaussian of width $3$ $\mu$m, to emulate the effect of the $3$ $\mu$m resolution. The fringe contrast is reduced as a result of the convolution, especially towards the back end of the pulse where the fringe separation is smaller. However, this effect is not enough to describe the lack of fringe contrast in the experimental images at the front of the pulse. One might suspect that the reduced fringe visibility is due to heating of the condensate atoms. However, no thermal fraction was observed. This was validated by repeating the experiment for different expansion times. Inspection of Figure \ref{expanded} shows that the wave front of the leading edge of the pulse is slightly curved. As this curvature will also be present in the direction of imaging, this will also contribute to the loss of fringe visibility.  Given these discrepancies, it is clear that the solutions of the GPE correctly capture the physical processes determining the evolution of the system.

\section{Conclusion}
We have generated shock waves in a large number elongated Bose-Einstein condensate by suddenly
splitting it with a blue-detuned optical dipole potential.  We observe no significant particle loss
or heating over these time scales, and find excellent agreement with the predictions of our
simulations of the 1D non-polynomial Gross-Pitaevskii equation.  We used a method to describe the
expansion of the elongated condensate while simulating the continuing dynamics in the long
direction, and again find excellent agreement with simulations.  Our study provides further
evidence that the Gross-Pitaevskii equation can be an excellent approximation to the dynamics of
condensates even in situations exhibiting violent dynamics and low dissipation.

\section{Acknowledgments}
This work was supported by the Australian Research Council Centre of Excellence for Quantum-Atom Optics and by Australian Research Council discovery project DP0985142 and the Stichting voor Fundamenteel Onderzoek der Materie (FOM) . We would like to acknowledge useful discussions with Peter Engels.  S. A. Haine would also like to acknowledge useful discussion with Andy Ferris.


\begin{thebibliography}{99}
\bibitem{anderson} M.H. Anderson, J.R. Ensher, M.R. Matthews, C.E. Wieman, and E.A. Cornell, Science {\bf 269}, 198 (1995).
\bibitem{gpe} Franco Dalfovo, Stefano Giorgini, Lev P. Pitaevskii, and Sandro Stringari, Rev. Mod. Phys. {\bf 71}, 463 (1999).
\bibitem{edwards} M. Edwards, P. A. Ruprecht, K. Burnett, R. J. Dod, and C. W. Clark, Phys. Rev. Lett. {\bf 77}, 1671 (1996).

\bibitem{burger} S. Burger, K. Bongs, S. Dettmer, W. Ertmer, K. Sengstock, A. Sanpera, G. V. Shlyapnikov, and M. Lewenstein, Phys. Rev. Lett. {\bf 83}, 5198 (1999). 

\bibitem{trippenbach} M. Trippenbach, Y. B. Band, and P. S. Julienne, Phys. Rev. A {\bf 62} 023608 (2000).

\bibitem{dutton} Z.~Dutton, M.~Budde, C.~Slowe and L.~V.~Hau, Science {\bf 293}, 663 (2001).

\bibitem{truscott} V. Y. F. Leung, A. G. Truscott, and K. G. H. Baldwin, Phys. Rev. A, 66, 061602 (2002). 

\bibitem{katz} N. Katz, R. Ozeri, J. Steinhauer, N. Davidson, C. Tozzo, and F. Dalfovo, Phys. Rev. Lett. {\bf 93}, 220403 (2004).

\bibitem{robins_rf1} N. P. Robins, C. M. Savage, J. J. Hope, J. E. Lye, C. S. Fletcher, S. A. Haine, and J. D. Close, , Phys. Rev. 4 {\bf 69}, 051602 (2004).

\bibitem{robins_rf2} N. P. Robins, A. K. Morrison, J. J. Hope, and J. D. Close, Phys. Rev. 4 {\bf 72}, 031606 (2005).

\bibitem{kohl} M. Kohl, Th. Busch, K. Molmer, T. W. Hansch, and T. Esslinger, Phys. Rev. A {\bf 72}, 063618 (2005). 

\bibitem{lye} J. E. Lye, L. Fallani, M. Modgno, D. S. Wiersma, C. Fort, and M. Inguscio, Phys. Rev. Lett. {95} 070401 (2005).

\bibitem{gpe_interferometry} T. Ananikian and T. Bergeman, Phys. Rev. A {\bf 73}, 013604 (2006).
\bibitem{pethick} C. J. Pethick and H. Smith. Bose-Einstein Condensation in Dilute Gases (Cambridge University Press, 2001).
\bibitem{norrie} A. A. Norrie, R. J. Ballagh, and C. W. Gardiner, Phys. Rev. A {\bf 73} 043617 (2004).
\bibitem{damski} B. Damski, Phys. Rev. A {\bf 69}, 043610 (2004).
\bibitem{Donley01}
E.~A. Donley, N.~R. Claussen, S.~L. Cornish, J.~L. Roberts, E.~A. Cornell and
  C.~E. Wieman; Nature \textbf{412}, 295 (2001).

\bibitem{Saito02} H. Saito and M. Ueda, Phys. Rev. A \textbf{65}, 033624 (2002).

\bibitem{Adhikari02} S. K. Adhikari, Phys. Lett. A, \textbf{296}, 145 (2002).

\bibitem{Santos02} L. Santos and G. V. Shlyapnikov, Phys. Rev. A \textbf{66},  011602   (2002).

\bibitem{Savage03}
C.~M. Savage, N.~P. Robins and J.~J. Hope, Phys. Rev. A
  \textbf{67}, 014304 (2003).

\bibitem{Wuester05}
S.~W{\"u}ster, J.~J. Hope and C.~M. Savage, Phys. Rev. A  \textbf{71}, 033604 (2005).

\bibitem{Wuester07} S. W\"{u}ster, B. J. Dabrowska-W\"{u}ster,  A. S. Bradley, M. J. Davis,
P. B. Blakie, J. J. Hope, and C. M. Savage,
Phys. Rev. A \textbf{75}, 043611 (2007).

\bibitem{Cornish06}
S.~L. Cornish, S.~T. Thompson and C.~E. Wieman, Phys. Rev. Lett. \textbf{96}, 170401 (2006).

\bibitem{Beata08} B. J. Dabrowska-W\"{u}ster,  S. W\"{u}ster, and M. J. Davis,
arXiv:0812.0493v1


\bibitem{Strecker02}
K.~E. Strecker, G.~B. Partridge, A.~G. Truscott and R.~G. Hulet, Nature \textbf{417}, 150 (2002).
\bibitem{Khay02}
L.~Khaykovich, F.~Schreck, G.~Ferrari, T.~Bourdel, J.~Cubizolles, L.~D. Carr,
  Y.~Castin and C.~Salomon,
  Science \textbf{296}, 1290, (2002).

\bibitem{Leung03}
V.~Y.~F. Leung, A.~ G.~ Truscott and K.~G.~H. Baldwin, Phys. Rev. A \textbf{66}, 061202 (2002).

\bibitem{li_rev}
K.~E. Strecker, G.~B. Partridge, A.~G. Truscott and R.~G. Hulet, New J. Phys. \textbf{5},
  73 (2003).

\bibitem{stoof_solitons}
U.~A. Khawaja, H.~T.~C. Stoof, R.~G. Hulet, K.~E. Strecker and G.~B. Partridge,
Phys. Rev.  Lett. \textbf{89}, 200404 (2002).


\bibitem{Simula05}
T. P. Simula, P. Engels, I. Coddington, V. Schweikhard, E. A. Cornell and R. J. Ballagh,
Phys. Rev. Lett. \textbf{94}, 080404 (2005).

\bibitem{hoefer} M.A. Hoefer, M.J. Ablowitz, I. Coddington, E.A. Cornell, P. Engels, and V.
Schweikhard, Phys. Rev. A {\bf 74}, 023623 (2006).
\bibitem{hoefer2} M.A. Hoefer, M.J. Ablowitz, P. Engels, Phys. Rev. Lett. {\bf 100}, 084504 (2008).

\bibitem{Chang08} J.~J.~Chang, P.~Engels and M.~A.~Hoefer, Phys. Rev. Lett. {\bf 101} 170404 (2008).

\bibitem{morgan} S. A. Morgan, R. J. Ballagh, and K. Burnett, Phys. Rev. A {\bf 55}, 4338 (1997)
\bibitem{stam} K.M.R. van der Stam, E.D. van Ooijen, R. Meppelink, J.M. Vogels, and P. van der Straten, Rev. Sci. Instrum. {\bf 78}, 013102
(2007).
\bibitem{andrews} M. R. Andrews, M.-O. Mewes, N. J. van Druten, D. S. Durfee, D. M. Kurn, W. Ketterle, Science {\bf 273},
84 (1996).
\bibitem{salasnich} L. Salasnich, A. Parola, and L. Reatto, Phys. Rev. A {\bf 65} 043614 (2002).
\bibitem{castin_dum} Y. Castin and R. Dum, Phys. Rev. Lett. {\bf 77}, 5315 (1996).

\end{thebibliography}
\end{document}